\pdfoutput=1

\documentclass[fleqn,usenatbib]{mnras}

\usepackage[T1]{fontenc}

\DeclareRobustCommand{\VAN}[3]{#2}
\let\VANthebibliography\thebibliography
\def\thebibliography{\DeclareRobustCommand{\VAN}[3]{##3}\VANthebibliography}

\usepackage{color}

\usepackage{graphicx}   
\usepackage{amsmath}    
\usepackage{amssymb}    

\title[Topological ignition of the stealth CME]{
Topological ignition of the stealth coronal mass ejections}

\author[Yu. V. Dumin \& B. V. Somov]{
Yurii V. Dumin,$^{1}$%
\thanks{E-mail: \href{mailto:dumin@yahoo.com}{dumin@yahoo.com} (YVD)}
and Boris V. Somov\thanks{Deceased}
\\
$^{1}$Sternberg Astronomical Institute of Lomonosov Moscow State University,
Universitetskii prosp.\ 13, Moscow, 119234 Russia}

\date{Received 2025 April 9, revised 2025 July 11}

\pubyear{2025}

\begin{document}
\label{firstpage}
\pagerange{\pageref{firstpage}--\pageref{lastpage}}
\maketitle

\begin{abstract}
One of hot topics in the solar physics are the so-called `stealth' coronal
mass ejections (CME), which are not associated with any appreciable energy
release events in the lower corona, such as the solar flares.
It is often assumed recently that these phenomena might be produced by some
specific physical mechanism, but no particular suggestions were put forward.
It is the aim of the present paper to show that a promising explanation of
the stealth CMEs can be based on the so-called `topological' ignition of
the magnetic reconnection, when the magnetic null point is produced by
a specific superposition of the remote sources (sunspots) rather than by
the local current systems.
As follows from our numerical simulations, the topological model explains
very well all basic features of the stealth CMEs:
(i)~the plasma eruption develops without an appreciable heat release
from the spot of reconnection, i.e., without the solar flare;
(ii)~the spot of reconnection (magnetic null point) can be formed far away
from the location of the magnetic field sources;
(iii)~the trajectories of eruption are usually strongly curved, which
can explain observability of CMEs generated behind the solar limb.
\end{abstract}



\section{Introduction}
\label{sec:Intro}

Since the discovery of the coronal mass ejections (CME) in the early 1970's
\citep[e.g. review][and references therein]{Howard_06}, it was
known that in some cases they could be reliably associated with other
manifestations of the solar activity in the lower corona (first of all,
the solar flares); while in other cases it was impossible to trace such
a relationship~\citep{Chen_11,Webb_12,Howard_13,Nitta_21,Reva_24}.
However, the undetectable origin of the respective CMEs was attributed for
a long time just to the insufficient quality of observations.

Meanwhile, in the course of development of the observational technique it was
gradually recognized that the initiation of some CMEs might be inherently
unobservable.
As a result, the term `stealth CME' emerged and became widely used in
the last decade.
Unfortunately, the nature of these phenomena remains unclear till now.
One point of view is that there is a continuous spectrum of CMEs with
various expression of energy release in the lower corona; and the stealth
CMEs belong just to one of the wings of this spectrum.
Another point of view is that there should be a special mechanism for
the production of such CMEs, but its physical principles are still unknown.

It is the aim of the present paper to show that a promising candidate
for the above-mentioned mechanism is the so-called `topological' ignition
(or trigger) of the magnetic reconnection, whose general principles were
formulated quite a long time ago by \citet{Gorbachev_88a} but remained
poorly exploited till now.
Here, we perform the detailed numerical simulations of the respective
process and show that its basic features are in perfect agreement with
the observed properties of the stealth CMEs.

\section{Theoretical model and simulations}
\label{sec:Model}

\begin{table*}
\caption{Basic features of the `standard' and `topological' mechanisms
of the magnetic reconnection.}
\label{Table_1}
\centering
\begin{tabular}{rlllll}
\hline\hline \\[-1.7ex]
&&& \multicolumn{1}{c}{\bf Standard reconnection} &&
    \multicolumn{1}{c}{\bf Topological reconnection} \\[0.6ex]
\hline \\[-1.7ex]
1. &   Source of the magnetic field &&
       Local electric currents &&
       Superposition of the remote sources \\[0.7ex]
2. &   Geometrical structure of the &&
       Determined by the local &&
       Determined by the global configuration \\
   &   reconnecting region &&
       magnetic field lines &&
       of the magnetic field \\[0.7ex]
3. &   Speed of propagation of the &&
       Limited by the Alfvenic &&
       Irrelevant to the Alfvenic velocity \\
   &   reconnection in space &&
       velocity \\[0.7ex]
4. &   Heat release in the spot of &&
       Substantial &&
       Insignificant \\
   &   reconnection &&
       && \\[0.7ex]
\hline
\end{tabular}
\end{table*}

It is commonly believed that the main source of the solar activity is
the magnetic reconnection, when the magnetic field lines break apart and
then connect again in a new configuration
\citep[e.g. monographs][]{Priest_00,Somov_13}.
This process takes place in the so-called null (or neutral) points, where all
components of the magnetic field vanish \citep{Parnell_96,Dumin_16}.
Such a null point is usually assumed to be formed by \textit{the local current
systems}; and this corresponds to the `standard' scenario of magnetic
reconnection.
On the other hand, there is yet another option for the appearance of the null
point, which is often overlooked.
This is a specific superposition of influences by \textit{the distant sources},
which can result in the X-type configuration with vanishing magnetic field
in its centre.
The possibility of such superposition was proved for the first time by
\citet{Gorbachev_88a} by utilizing rather sophisticated theorems of
differential geometry and algebraic topology, and the respective effect was
called the `topological trigger' of magnetic reconnection.
The most important differences between the `standard' and `topological'
scenarios are summarized in Table~\ref{Table_1}.

Our consideration will be based on the above-mentioned
Gorbachev--Kel'ner--Somov--Shvarts (GKSS) model, whose specific feature is
the existence of the `topologically unstable' arrangements of
the magnetic-field sources (sunspots).
Their tiny variation results in the dramatic reconstruction of the magnetic
field in the entire space; and just this effect will be substantially
employed below \citep[for a particular example of the unstable
configuration, see Fig.~2 in][]{Dumin_24}.

A few topological models of another type were developed in the late 1990's
and early 2000's by the group of E.R.~Priest
\citep[for example,][]{Brown_99,Inverarity_99,Brown_01}.
They assumed a rather regular arrangement of the sources in
the photospheric plane and then considered the emergence of a new null
point as a result of its `squashing' out of this plane, when the
above-mentioned sources converged.
Unfortunately, eruption of the null point in such models turns out to
be rather slow (of the same order as velocity of the sources);
and, therefore, they are less appropriate for the description of CMEs.
\citep[For application of the topological methods to various solar
phenomena, see review][and references therein.]{Longcope_05}

\begin{figure}
\centering
\includegraphics[width=0.9\hsize]{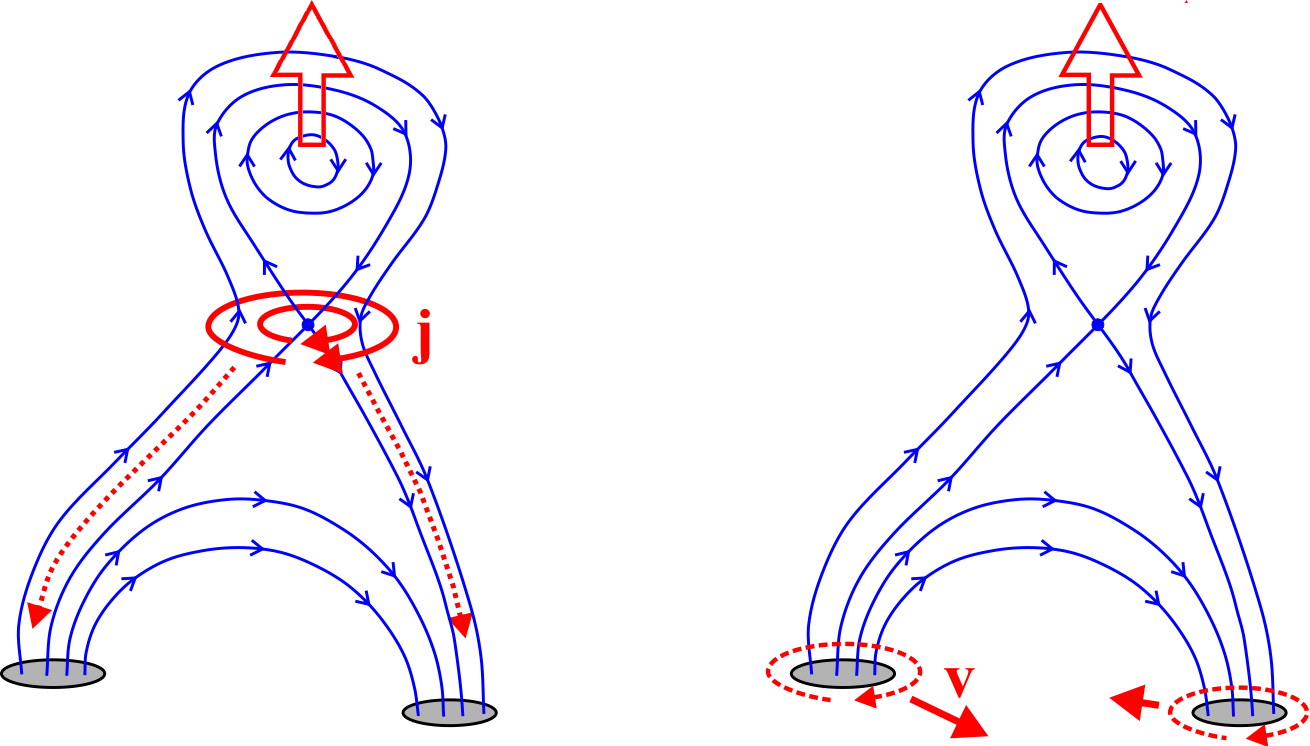}
\caption{Sketch of development of the magnetic reconnection in
the standard (left panel) vs.\ the topological (right panel) scenarios.}
\label{Fig1}
\end{figure}

The first~-- and most important~-- reason why the topological models look
very promising for the description of stealth CMEs is illustrated in
Fig.~\ref{Fig1}:
In the `standard' scenario, development of the magnetic reconnection
leads both to a heat release due to the dissipation of the local
current system~{\bf j} (resulting in the solar flare) and a detachment
of the plasma bunch from the null point (the CME eruption).
On the other hand, in the `topological' scenario, the reconnection
is caused by the motions~{\bf v} of the photospheric sources, without
any currents immediately at the spot of reconnection.
This results solely in the CME eruption.

\begin{figure*}
\centering
\includegraphics[width=0.82\hsize]{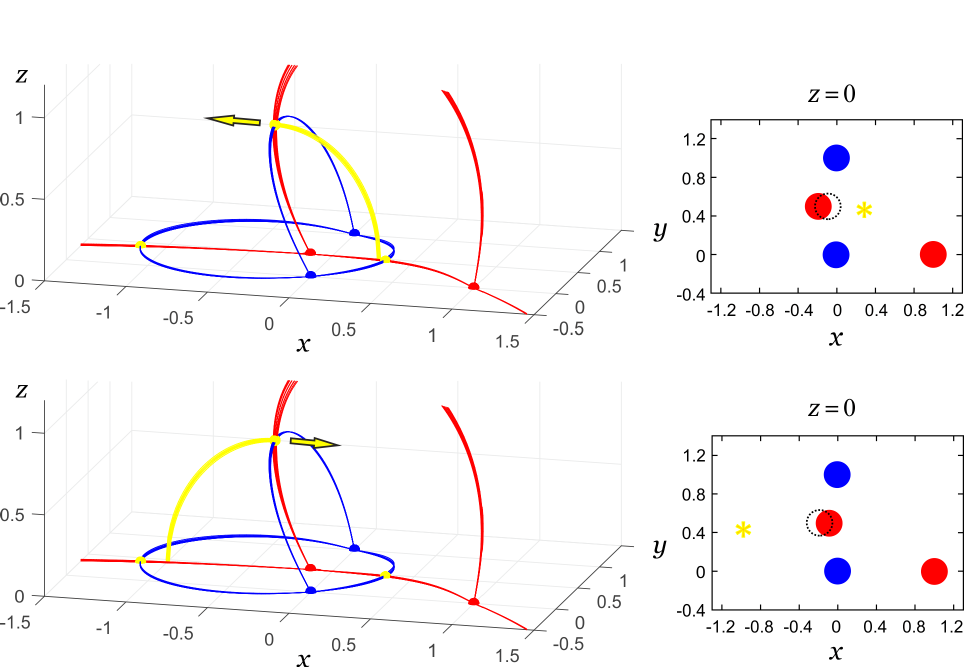}
\caption{Left panels: magnetic skeletons (red and blue curves) and
trajectories of eruption of the null points (yellow curves)
in the region of topological instability.
Right panels: respective arrangements of the magnetic sources
in the plane of photosphere before (solid circles) and after
(dotted circumferences) development of the instability.
The spots of emergence of the null points (i.e.,
onset of the eruption) are designated by the yellow stars.
The corresponding animated images are available as movies in
the Supplementary material.}
\label{Fig2}
\end{figure*}

Moreover, apart from the above-mentioned `energetic' argument, there
are two additional `geometric' arguments in favor of the topological
mechanism as explanation of the stealth CMEs.
They follow from the simulations presented in Fig.~\ref{Fig2}.
The magnetic field was assumed to be formed by the two pairs of
the point-like sources of equal magnitude but opposite signs located
in the plane of photosphere or somewhat below it.
From the physical point of view, they represent open ends of
magnetic-flux tubes originating in the deeper layers of the Sun
\citep[for an additional discussion, see the introductory section
in][]{Zhuzhoma_22}.
The global magnetic field in this situation is described by
the so-called `two-dome structure', which separates the entire space
into the four topologically distinct subregions
\citep[e.g. Fig.~3 in][]{Somov_08}.
Some further mathematical details of such simulations can be found
in Appendix~A of paper by \citet{Dumin_19} and
Appendix~\ref{sec:Math_form} of the present paper.

As was predicted by \citet{Gorbachev_88a}, there are specific
`topologically unstable' arrangements of the magnetic sources, when
a tiny shift of one of them results in a dramatic reconstruction of
the entire magnetic field.
The most well-known arrangement of the unstable type is formed when
the sources are located approximately in the vertices of the slanted
letter {\it `T'}, as illustrated in Fig.~\ref{Fig4}; see also right panel
of Fig.~2 in paper by \citet{Dumin_24}.
(Subsequently, configurations of the same type, but composed of two dipoles,
were studied in much more detail by \citet{Beveridge_02}.)
So, if the unstable configuration is realized, the above-mentioned
two-dome structure experiences a sudden flipping, as is seen in
the left panels of Fig.~\ref{Fig2} and, especially, in
Supplementary movies.
To avoid cluttering the pictures with unnecessary details,
we represent only the `topological skeletons', which are the sets
of magnetic-field lines connecting the sources and null points;
two different colors (red and blue) refer to the field lines
emanating from the sources of two different polarities.

Next, which is most important, the above-mentioned flipping
leads to emergence and fast motion of a new null point high
above the plane of sources, as shown by the yellow curves.
Therefore, a magnetic reconnection~-- unrelated to any local
currents~-- should develop along this trajectory; and a plasma blob
can detach and escape away somewhere at this curve.
Just this effect might be a reasonable explanation for
the formation of stealth CMEs, since it is not associated with
any heat release.

\begin{figure}
\centering
\includegraphics[width=0.7\hsize]{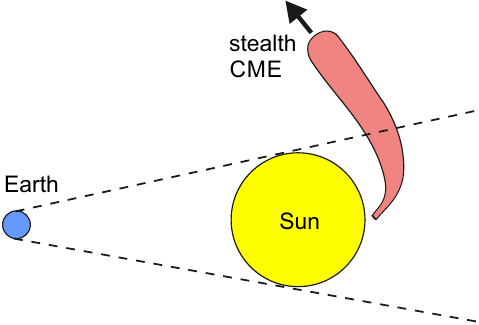}
\caption{Sketch of the strongly-bended trajectory of CME, facilitating
its observation from the Earth.}
\label{Fig3}
\end{figure}

One can also see in the simulations two important geometrical properties
of the null-point trajectories:
\begin{enumerate}
\item
They are strongly bended.
Therefore, even if such an eruption occurred on the back side of
the Sun, it might be well observable from the Earth, as illustrated
in Fig.~\ref{Fig3}.
\item
Directions of the outbursts are crucially dependent on the particular
type of motion of the sources in the region of topological instability.
Namely, if the central source is shifted to the right, the eruption
originates somewhere inside the source region and is directed outwards
(two upper panels in Fig.~\ref{Fig2}).
On the other hand, if the central source is shifted (almost from
the same position) to the left, then the eruption originates quite
far away from the source region and is directed towards this region.
Such a behaviour is illustrated more pictorially in Fig.~\ref{Fig4}.
Therefore, approximately in 50~per cent of cases the eruption looks
`detached' from the sources (i.e., the active region on the Sun).
\end{enumerate}

\begin{figure}
\centering
\includegraphics[width=0.6\hsize]{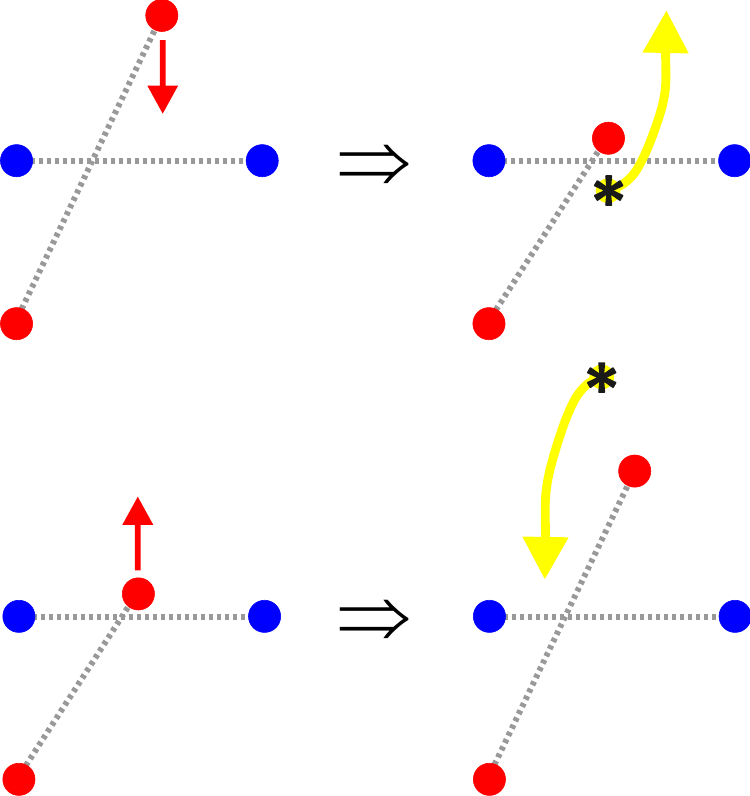}
\caption{Two types of eruption of the null point:
(top panel) the outward eruption, when X-type configuration transforms
into the T-type one; and (bottom panel) the inward eruption, when T-type
configuration transforms into the X-type one.
Blue and red circles are the point-like magnetic sources of opposite
signs, and the straight arrows show the directions of their motion.
Yellow curves with arrows designate trajectories of the null points, and
the spots of their origin are marked by black stars.
The dashed grey lines are drown just for a visual aid.
(These pictures are rotated by~$ 90^{\rm o} $ with respect to the right
panels in Fig.~\ref{Fig2}.)}
\label{Fig4}
\end{figure}

\section{Discussion}
\label{sec:Discussion}

As follows from the above consideration, there are three
arguments~-- one energetic and two geometrical~-- why the topological
mechanism is a promising explanation of the stealth CMEs;
and the observational findings are in reasonable agreement with our
simulations.

Really, as was mentioned by~\citet{Nitta_21}, the stealth CMEs are
faint and slow, which assumes that they involve less energy than normal
CMEs.
However, this feature was attributed by the above-cited authors just to
the empirical fact that the magnetic energy available for an eruption
above a quiet Sun or decayed active region should be substantially less
than the energy from a volume above a strong-field active region.
On the other hand, the topological model suggests a much more profound
basis for explanation of the low energy.

Besides, as emphasized by~\citet{Howard_13}, ``CMEs themselves can erupt
with an invisible or almost invisible signature in coronagraphs as well'',
because they may be just the ``erupting magnetic structures'' and
``take place without containing sufficient excess plasma.''
In fact, just this property is postulated in the formulation of
our topological model.

Next, the first geometric feature~-- namely, a strongly-curved trajectory
of ejection~-- can answer the question posed, e.g., by \citet{Robbrecht_09}:
how can a CME originating on the back side of Sun have any geoeffective
action?
The second geometric feature~-- an apparent `detachment' of the spot of
eruption from the region of magnetic sources, which is realized in
about half of the cases~-- can explain the so-called `random' coincidences
of the CMEs with the strong flares in the opposite quadrant of the solar
disc \citep[e.g. Fig.~3(d) in paper][]{Reva_24}.

Finally, let us mention that over a decade ago \citet{Oreshina_12} already
speculated that formation of CMEs could be caused by the topological
instability (or `topological trigger') of the coronal magnetic fields,
but they discussed only the CMEs associated with solar flares.
Unfortunately, it was overlooked that even a more promising scope of
applicability of the topological models might be just the stealth CMEs.

\section{Conclusions}
\label{sec:Conclusions}

The topological models of magnetic reconnection were suggested over
35~years ago but were repeatedly criticized since that time for their
inability to explain an appreciable heat release due to the reconnection.
However, this disadvantage transforms into the crucial advantage when
it is necessary to interpret the stealth coronal mass ejections,
occurring without the solar flares.
Really, the topological mechanism can easily explain a global
reconstruction of the magnetic field without any heat release in
the vicinity of the null point responsible for the reconnection.
Moreover, the geometrical properties of eruptions characteristic of
the topological models (namely, their strongly curved trajectories and
a `detachment' of the spot of eruption from the source region) are
also very favorable for explanation of the stealth CMEs.

At last, it is important to emphasize that the topological models
do not require to introduce any kind of the `new physics'.
In fact, they are based just on the thoroughout analysis of
the specific magnetic-field configurations on the basis of Maxwell
equations, which are even simpler than the standard MHD.
Of course, a more realistic simulation of the stealth CMEs requires
a detailed self-consistent MHD modeling, starting from the configurations
outlined here.


\section*{Acknowledgements}

YVD is grateful to
P.M.~Akhmet'ev,
B.P.~Filippov,
V.V.~Kalegaev,
V.D~Kuznetsov,
A.T.~Lukashenko,
A.V.~Oreshina,
I.~Slezak (former I.V.~Oreshina),
D.D.~Sokoloff,
N.A.~Vlasova,
and
E.V.~Zhuzhoma
for valuable discussions and consultations.
The study was conducted under the state assignment of Lomonosov Moscow
State University.

Authors' contributions:
BVS suggested the concept of `topological' magnetic reconnection,
supervised the work and discussed the results.
YVD suggested to apply the topological reconnection to interpret
stealth CMEs, performed the corresponding numerical simulations
and prepared the manuscript.


\section*{Data Availability}

Computer software used for the simulations can be obtained from
YVD by reasonable request.


\bibliographystyle{mnras}


\begin{appendix}

\section{Basic mathematical formalism}
\label{sec:Math_form}

Analysis in the framework of topological models is usually performed in
two steps:
First of all, structure of the magnetic field~$ \bf B $ is calculated in
the magnetostatic approximation:
\begin{equation}
\mathbfit{B} = -\nabla \Phi \, ,
\qquad
\Delta \Phi = 0 \, ,
\label{eq:Pot_field}
\end{equation}
where $ \Phi $~is the scalar magnetic potential.
So, the time enters into solution only as a parameter, due to the temporal
dependence of the magnetic sources (which are usually assumed to be located
at the boundary of the volume under consideration, e.g., at the level
of photosphere or somewhat below it).
Next, the plasma motion and the associated processes (heating, emission,
etc.) are analysed in the given magnetic field.

A key point in the construction of topological model is searching for
the specific `topologically unstable' arrangement of the sources, where
their small variation results in the emergence of a new null point and its
subsequent fast eruption out of the plane of the sources.
As seen in the simulations presented in Section~\ref{sec:Model}, this
process is associated with sharp reconstruction of the magnetic field in
the entire space.

The general mathematical criteria of the topological instability were
formulated by \citet{Gorbachev_88a} as
\begin{equation}
\left\{
\begin{aligned}
& \, \mathbfit{B}(\mathbfit{r})%
  \Big|_{\mathbfit{r}=\mathbfit{r}^*} \!\! = 0 \, , \\
& \, \det\big|\!\big| \,
  {\partial}^2 \Phi \big/ {\partial r_i}{\partial r_j} \,
  \big|\!\big| \, \Big|_{\mathbfit{r}=\mathbfit{r}^*} \!\! = 0 \, .
\end{aligned}
\right.
\label{eq:Topol_instab}
\end{equation}
In other words, it is necessary to find the arrangement of magnetic
sources admitting the null points~$ \mathbfit{r}^* $ in which the
Euler--Poincar\'{e} topological indices change their signs.

In general, searching for the configurations satisfying
equation~(\ref{eq:Topol_instab}) is a very hard mathematical task.
By now, the corresponding analysis was performed only in the approximation
of point-like sources (the effective `magnetic charges'), when
the magnetic field can be presented as
\begin{equation}
\Phi(\mathbfit{r}) = \sum\limits_i e_i \,
  \frac{\mathbfit{r} - \mathbfit{r}_i}{|\mathbfit{r} - \mathbfit{r}_i|^3} \, ;
\label{eq:Phi_4_charges}
\end{equation}
for the additional details, see papers by \citet{Gorbachev_88a} and
\citet{Somov_08}.

\begin{figure}
\centering
\includegraphics[width=0.82\hsize]{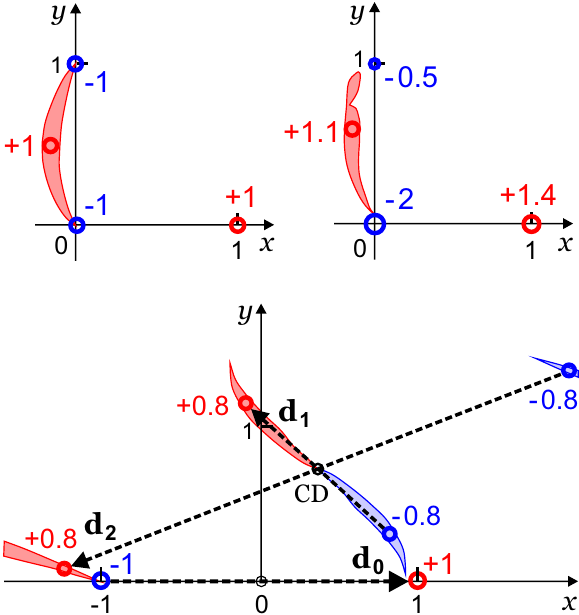}
\caption{
Examples of the magnetic source arrangements exhibiting the
`topological instability', i.e. formation of the `coronal null':
(top panels) \citet{Gorbachev_88a} model, where three magnetic sources
are fixed in the vertices of a triangle, while the fourth source can take
an arbitrary position;
and (bottom panel) \citet{Beveridge_02} model, composed of two
dipoles~-- either $ {\bf d}_0 $ and $ {\bf d}_1 $ or
$ {\bf d}_0 $ and $ {\bf d}_2 $~-- with a fixed center~CD.
}
\label{Fig5}
\end{figure}

Figure~\ref{Fig5} illustrates two groups of the topologically unstable
arrangements, which were studied by now in most detail.
The first of them is the T-type configuration formed by two pairs of
sources of the opposite polarity \citep{Gorbachev_88a}.
Just this arrangement was used in our numerical simulations.
The second arrangement is composed of two magnetic dipoles.
One of them is fixed ($ {\bf d}_0 $), while another ($ {\bf d}_1 $ or
$ {\bf d}_2 $) can change its orientation and length about a fixed
center~CD.
The regions of instability are shaded in red (for the positive source)
and blue (for the negative one).
To avoid misunderstanding, let us emphasize that~-- from the viewpoint
of rigorous mathematical terminology~-- the criteria of instability
are sufficient rather than necessary ones.

\end{appendix}

\bsp    
\label{lastpage}

\end{document}